\documentclass[preprint,times]{elsarticle}
\pdfoutput=1
\usepackage{amsmath}  
\usepackage{amssymb}
\usepackage{subfigure}
\usepackage{ecrc}
\usepackage{rotating}
\usepackage{xcolor}
\usepackage{latexsym}
\usepackage{bm}
\usepackage{floatrow}
\usepackage{breqn}
\newfloatcommand{capbtabbox}{table}[][0.45\textwidth]
\usepackage{tikz}
\newcommand*\circled[1]{\tikz[baseline=(char.base)]{
            \node[shape=circle,draw,inner sep=0.5pt] (char) {\footnotesize #1};}}
\volume{}

\jid{}
\firstpage{1}
\journalname{}
\jnltitlelogo{}
\runauth{}

\begin{document}
\begin{frontmatter}

\title{Theoretical estimates for flat voids coalescence by internal necking}

\author[CEA]{J.~Hure\corref{cor1}}
\author[CEA]{P.O.~Barrioz}

\cortext[cor1]{Corresponding author}
\address[CEA]{CEA Saclay, Universit\'e Paris-Saclay, DEN, Service d'\'Etudes des Mat\'eriaux Irradi\'es, 91191 Gif-sur-Yvette, France}

\begin{abstract}

Coalescence of voids by internal necking is in most cases the last microscopic event related to ductile fracture and corresponds to a localized plastic flow between adjacent voids. Macroscopic load associated to the onset of coalescence is classically estimated based on limit analysis. However, a rigorous upper-bound mathematical expression for the limit-load required for flat voids coalescence  that remains finite for penny-shaped voids/cracks is still unavailable. Therefore, based on limit analysis, theoretical upper-bound estimates - both integral expression and closed-form formula - are obtained for the limit-load of cylindrical flat voids in cylindrical unit-cell subjected to boundary conditions allowing the assessment of coalescence, \textcolor{black}{for axisymmetric stress state}. These estimates, leading to finite limit-loads for penny-shaped cracks, are shown to be in very good agreement with numerical limit analysis, for both cylindrical and spheroidal voids. \textcolor{black}{Approximate formula is also given for coalescence under combined tension and shear loading. These coalescence criteria} can thus be used to predict onset of coalescence of voids by internal necking in ductile fracture modelling. 
\end{abstract}

\begin{keyword}
Ductile fracture, Voids, Coalescence, Necking, Limit analysis, Homogeneization
\end{keyword}

\end{frontmatter}

\section{Introduction}

Macroscopic ductile fracture of metallic materials has been shown to be related to three different microscopic phenomena, namely void initiation, growth and coalescence \cite{puttick}. In most cases, debonding or cracking of second-phase particles lead to the creation of voids \cite{argon75,beremin81,babout04}, that grow under mechanical loading \cite{ricetracey,mcclintock,marini,weck} until localization appears between adjacent voids leading to void coalescence \cite{brown,thomason68,cox,wecktoda}. Homogeneized macroscopic models of porous materials have been developed since the work of Gurson \cite{gurson} based on limit analysis and Rousselier \cite{rousselier} based on thermodynamical concepts, and more recently based on non-linear variational principles \cite{danas}. Models have been extended to account for void nucleation \cite{chu}, void shape effects \cite{gologanu,madou0}, plastic anisotropy effects (with or without void effects) \cite{benzergaanisotrope,benzergagld,monchiet,morinellipse} and more recently to single crystals \cite{xuhan,paux,mbiakop}. Both microscopic and homogeneized macroscopic models have been assessed through comparisons to computational cell models (see, \textit{e.g.}, \cite{koplik,tvergaard}). Homogeneized models attempt to account for void coalescence, either by considering an empirical acceleration of porosity after a critical value \cite{tvergaardneedleman} or by coupling directly growth model to coalescence model, the latter giving a flow potential after the onset of coalescence \cite{benzergagrowthcoalescence}. The last method appears to be the most promising as no empirical parameters identification is necessary, but requires accurate void coalescence model which is the main purpose of the present work and will be detailed hereafter. The interested reader is referred to the reviews of Benzerga and Leblond \cite{benzergaleblond}, Besson \cite{besson2010} \textcolor{black}{and Pineau \textit{et al.} \cite{pineaureview}} for detailed presentation about ductile fracture.

Experimental observations of void coalescence have distinguished three main mechanisms: \textit{internal necking} where localized plastic flow appears in the intervoid ligament perpendicular to the main loading direction similar to necking observed in tensile test, \textit{void sheeting} involving shear band \cite{cox}, and \textit{necklace coalescence} where localized plastic flow appears in the intervoid ligament parallel to the main loading direction \cite{benzerganecklace}. In this study, we focus on void coalescence by internal necking, also refers to as coalescence in layers, which is the most common situation.
Thomason \cite{thomason68,thomason85a,thomason85b,thomason90} proposed to assess the \textit{onset of coalescence} through reaching plastic limit-load in the intervoid ligament, considering a perfectly plastic material and making use of both limit analysis and homogeneization technique. The analysis gives the macroscopic stress at which onset of coalescence - which can be seen as a transition from diffuse plastic flow around the void to localized flow in the intervoid ligament with associated elastic unloading in other regions - is expected to occur. Such kinematics have been supported by unit-cell simulations \cite{koplik}. Thomason model have been shown to give quite accurate predictions compared to experimental results (see for example \cite{wecktoda}), and have been extended to account for strain hardening \cite{pardoen,scheyvaerts} and secondary voids \cite{fabregue}.

\begin{figure}[H]
\centering
\includegraphics[height = 4cm]{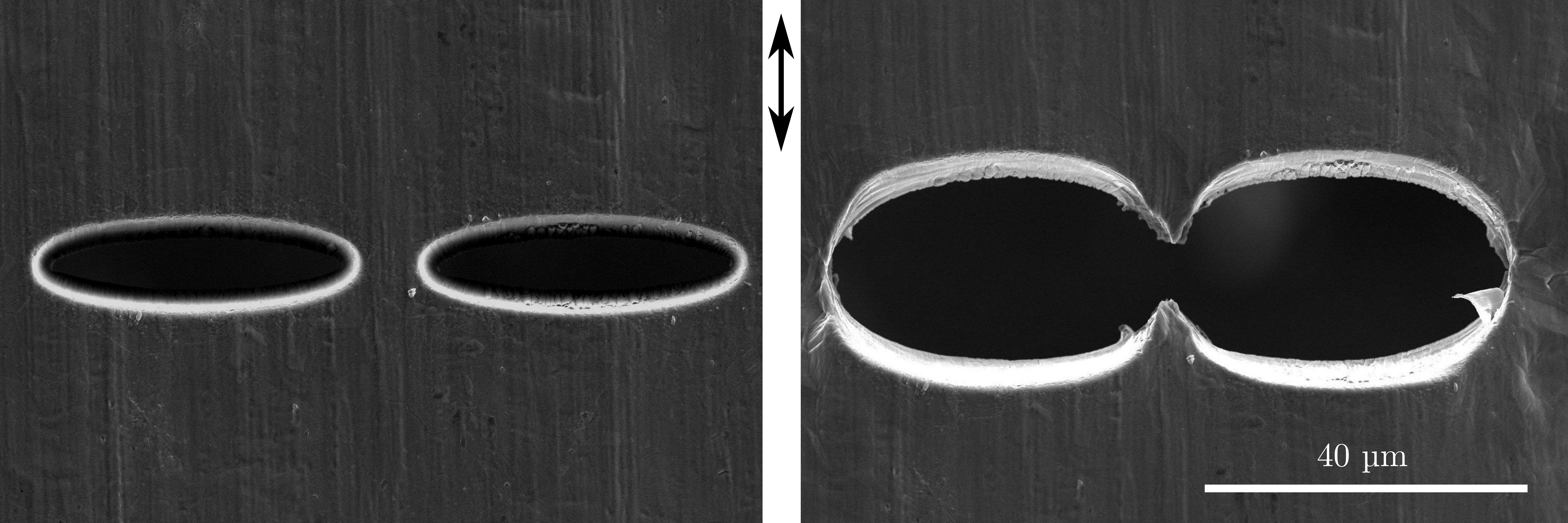}
\caption{\textcolor{black}{Experimental observation of flat void coalescence by internal necking. Ellipsoidal holes have been drilled through $75\mu$m stainless steel sheet and subjected to uniaxial tension. Black arrow indicates the loading direction.}}
\label{fig0}
\end{figure}

The original Thomason model \cite{thomason90} - and extensions based on it - suffers from two main drawbacks. The first one is that it gives infinite coalescence load in the limit of very flat voids, \textit{i.e.} penny-shaped cracks. \textcolor{black}{Thus, predictions worsen as the void flatness increases.} This prevents for example the use of Thomason model in low stress triaxiality conditions, where initially spherical voids tend to form micro-cracks \cite{tvergaardshear}. \textcolor{black}{Moreover, although limited experimental results are yet available in literature, evidence of void coalescence of flat voids has been reported for aluminum alloys through Synchrotron Radiation Computed Tomography (SRCT) \cite{yangshen}. Voids initiate from elongated particles, and remain flat up to coalescence even though plastic deformation occurs because of the small distance between them. Example of flat void coalescence is shown in Fig.~\ref{fig0} on a model experiment. Aluminum alloys are used as structural materials in industrial applications, thus precise estimation of flat void coalescence is needed.} The second drawback is that, while the upper-bound theorem of limit analysis was used, Thomason relies at the end on an empirical equation that may not be strictly an upper-bound. Recently, rigorous theoretical upper-bound estimates were obtained by considering cylindrical voids in cylindrical unit-cell \cite{thomasonnew1,thomasonnew2}. \textcolor{black}{To account for finite limit-load of penny-shaped cracks, empirical modification of Thomason model has been proposed in \cite{benzerga02}, while heuristic modifications of rigorous models have been more recently proposed \cite{torki,keralavarma}}. However, to the knowledge of the authors, no mathematical expression is available that provides a rigorous upper-bound estimate of the limit-load that remains finite for penny-shaped cracks. Therefore, the aim of this paper is to provide rigorous upper-bound estimates of the limit-load for coalescence of flat voids by internal necking that leads to finite limit-load for penny-shaped cracks, in the case of axisymmetric loading conditions.\\

In the first part of the paper, cylindrical unit-cell with cylindrical and spheroidal void is described as well as the boundary conditions considered for the study of coalescence. Theoretical results of limit analysis for a von Mises matrix and numerical limit analysis are briefly summarized. In a second part, upper-bound estimates - both integral expression and closed-form formula - for the coalescence limit-load of cylindrical unit-cell containing flat cylindrical void \textcolor{black}{under axisymmetric loading conditions} are detailed and compared to supposedly exact numerical results. The estimates are finally compared to numerical results for flat spheroidal voids, and to the predictions of coalescence models with heuristic modifications to account for finite limit-load for penny-shaped cracks \cite{benzerga02,torki,keralavarma}. \textcolor{black}{In addition, approximate void coalescence criterion for combined tension and shear is proposed based on the coalescence limit-load obtained under axisymmetric loading conditions.}\\

\section{Limit analysis of cylindrical unit-cell with voids}

\subsection{Geometry and boundary conditions}

A cylindrical unit-cell $\Omega$ of half-height $H$ and radius $L$ containing a coaxial void $\omega$ is used in this study (Fig.~2). Two geometries of voids are considered, namely cylindrical (of radius $R$ and half-height $h$) and spheroidal (of semi-principal length $R$ and $h$). Two dimensionless ratio will be used in the following:
\begin{equation}
W = \frac{h}{R} \ \ \ \ \ \ \ \ \ \ \chi = \frac{R}{L}
\end{equation}
\noindent
where $W$ is the aspect ratio of the void, and $\chi$ the dimensionless length of the inter-void ligament. As only coalescence is studied here, \textit{i.e.} localized plastic flow in the inter-void ligament, the height $H$ is not a parameter of interest \cite{thomasonnew2}. The cylindrical unit-cell subjected to the following boundary conditions for the velocity field:
\begin{equation}
\begin{aligned}
v_r(L,z) &= D_{11}L\\
v_z(r,\pm H) &= \pm D_{33}H\\
\end{aligned}
\end{equation}

\noindent
stands as an approximation of a unit-cell of a periodic array of voids of hexagonal lattice\footnote{The reader is referred to \cite{kunasun} for a discussion about the effect of the choice of the unit-cell.} (Fig.~1) under periodic boundary conditions \cite{koplik} \textcolor{black}{in axisymmetric stress state ($\Sigma_{33} > \Sigma_{11} = \Sigma_{22}$, $\Sigma_{ij}=0\ \mathrm{for}\ i\neq j$).}

\begin{figure}[H]
\centering
\includegraphics[height = 7cm]{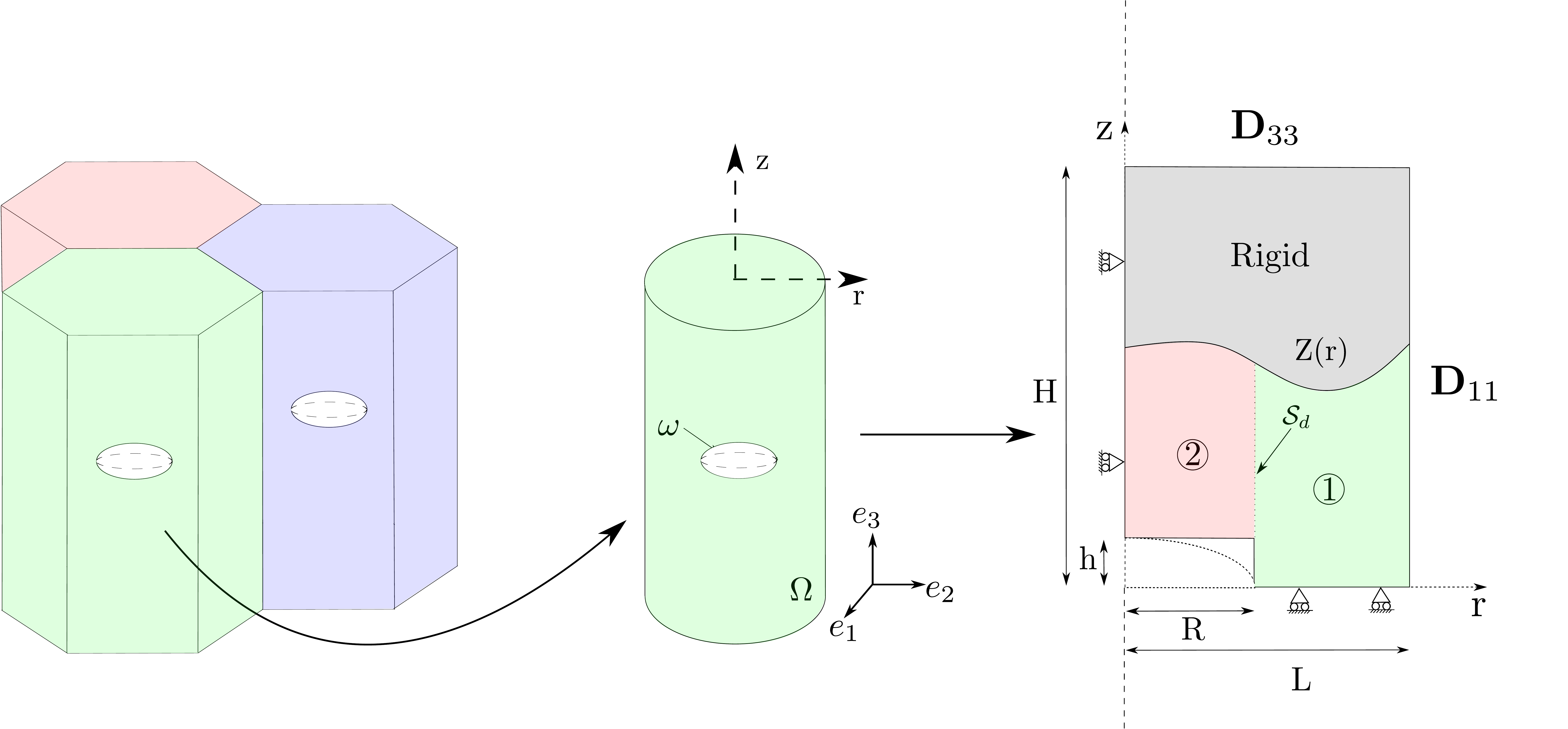}
\caption{Cylindrical unit-cell considered as an approximation of a unit-cell of a periodic array of voids of hexagonal lattice under periodic boundary conditions}
\end{figure}

The material is supposed to be rigid-perfectly plastic\footnote{As a classical result of limit analysis that will be used in this study is that elastic strain rates vanish at limit-load.}, obeying von Mises' criterion, and plastic flow is associated by normality. At coalescence, the regions above and below the void unload elastically \cite{koplik}, which thus corresponds in our modelling hypothesis to the presence of rigid regions. Finally, only half of the unit-cell (along the z-axis) will be considered because of symmetry. The boundary conditions for the velocity field in order to assess coalescence are:

\begin{equation}
\left\{
\begin{aligned}
v_r(0,z) = v_r(r,z\geq Z(r)) &= 0\\
v_z(r\geq R,0) &= 0\\
v_z(r,z\geq Z(r)) &= C^{ste}\\
\end{aligned}
\right.
\hspace{1cm}
\Rightarrow
\hspace{1cm}
\left\{
\begin{aligned}
D_{11} &= 0\\
D_{33} &= \frac{v_z(r,H)}{H}\\
\end{aligned}
\right.
\label{eqperiodic}
\end{equation}

\subsection{Theoretical estimates}

In order to evaluate the macroscopic stress at which onset of coalescence occurs, periodic homogeneization along with limit analysis is used (see, \textit{e.g.}, \cite{benzergaleblond} for details). For periodic boundary conditions, macroscopic stress $\Sigma$ and strain rate \textbf{D} tensors are related to their microscopic counterparts by volume averaging:
\begin{equation}
\Sigma = \frac{1}{vol{\Omega}} \int_{\Omega} \sigma d\Omega \ \ \ \ \ \ \ \ \ \ \textbf{D} = \frac{1}{vol{\Omega}} \int_{\Omega} \textbf{d} d\Omega
\label{macroscopic}
\end{equation}
with $\sigma$ the Cauchy stress and $\textbf{d}$ the microscopic strain rate tensor. Hill-Mandel lemma reads:
\begin{equation}
\frac{1}{vol{\Omega}} \int_{\Omega} \sigma:\textbf{d}\, d\Omega =  \Sigma:\textbf{D}
\end{equation}
Upper-bound theorem of limit analysis enables to assess the limit-load of the structure, and is, for a perfectly plastic material obeying von Mises' criterion:
\begin{equation}
\textcolor{black}{ \Sigma:\textbf{D} = \Pi(\textbf{D})  \leq \Pi^+(\textbf{D})}
\label{eqanalyselimite}
\end{equation}

\begin{equation}
\mathrm{with\ \ \ \ \ \  }\Pi^+(\textbf{D}) = \left< \sigma_0 d_{eq} \right>_{\Omega - \omega} = \frac{1}{vol{\Omega}} \int_{\Omega - \omega} \sigma_0 d_{eq} d\Omega 
\label{formulapi}
\end{equation}
where $\sigma_0$ is the yield stress, $d_{eq} = \sqrt{[2/3]\textbf{d}:\textbf{d}}$ the equivalent strain rate \textcolor{black}{($\textbf{d}=[\bm{^t\nabla} v + \bm{\nabla} v]/2)$}, \textcolor{black}{and $v$ a velocity field kinematically admissible with $\textbf{D}$ (Eqs.~\ref{eqperiodic}) and verifying the property of incompressibility $\mathrm{tr}(\textbf{d})=0$}. \textcolor{black}{$\Pi^+(\textbf{D})=\Pi(\textbf{D})$ when $v$ is the velocity field solution. $\Pi(\textbf{D})$ will be referred to as the macroscopic plastic dissipation and superscript $^+$ will be omitted in the following for clarity.} \textcolor{black}{In presence of a velocity field having a purely tangential discontinuity along an interface $S_d$, the plastic dissipation related to the discontinuity is:} 

\begin{equation}
\textcolor{black}{\Pi_{surf}(\textbf{D}) = \frac{1}{vol{\Omega}} \int_{Sd} \frac{\sigma_0}{\sqrt{3}} ||v_t||dS}
\label{formulapisurf}
\end{equation}
\textcolor{black}{where $||v_t||$ is} the absolute value of the velocity jump. The macroscopic limit stress or yield locus is obtained from Eq.~\ref{eqanalyselimite} by the equation:
\begin{equation}
\Sigma = \frac{\partial \Pi(\textbf{D}) }{\partial \textbf{D}}
\label{eqanalyselimite2}
\end{equation}
Analytical expression for macroscopic limit-load according to Eq.~\ref{eqanalyselimite2} that will stand as coalescence load requires the choice of trial velocity fields. 

\subsection{Numerical evaluation of limit-loads}
\label{num}
Limit analysis problem has been shown to be solved numerically efficiently with a standard finite element software by considering an elastic-perfectly plastic material, no geometry update (\text{i.e.} small strains approximation), implicit constitutive law integration,  and a single large loading step (large enough so that elastic strains are negligible compared to plastic strains) \cite{madou,tekoglu}. Under these hypotheses, the limit analysis problem is equivalent to the mechanical problem with $p \leftrightarrow d_{eq}$ and $v \leftrightarrow u$, with $p$ the cumulated plastic strain and $u$ the displacement at the end of the single loading step. Macroscopic limit-load can finally be computed by Eq.~\ref{macroscopic}, is exact, up to numerical errors, and can be compared to upper-bound estimates obtained through Eq.~\ref{eqanalyselimite2}.\\

Numerical evaluations of limit-load presented hereafter have been carried out with the finite element software Cast3M \cite{castem} under small strain assumption, with an elastic perfectly-plastic behavior (of Young's modulus $E$, Poisson ratio $\nu$ and yield stress $\sigma_0$). The ratio $E/\sigma_0$ was fixed to $10^4$, $\nu$ to 0.49 so that elastic stress prediction corresponds to an almost incompressible displacement field. Convergence of the mechanical equilibrium have been found to be difficult in most cases if one single loading step was applied. Therefore, multiple loading steps were applied to estimate the limit-load. \textcolor{black}{Boundary conditions defined in Eqs.~\ref{eqperiodic} are used, ensuring that a coalescence mode is obtained at the limit-load, \textit{i.e.} localized plastic flow in the intervoid ligament.} Comparison of theoretical trial velocity field and the associated equivalent strain rate can be made through the conversion $v \leftrightarrow \Delta u$ and $d_{eq} \leftrightarrow \Delta p$, where $\Delta u$ and $\Delta p$ are the increment of displacement and equivalent plastic strain over the last computation step, respectively. Mesh convergence was performed for all simulations results.

\section{Upper-bound estimates of coalescence load for flat voids: $W \leq 1$}

Theoretical upper-bound estimates for the coalescence limit-load of cylindrical flat voids are derived in this section. \textcolor{black}{Void coalescence is assumed to come from \textit{internal necking} where localized plastic flow occurs in the intervoid ligament. Plastic flow remains however diffuse with respect to the scales involved in the ligament (void radius and height, and intervoid distance)\footnote{\textcolor{black}{in opposition to \textit{void sheeting} coalescence involving localized plastic flow both with respect to the scales of the unit-cell and the scales of the ligament}}. Even in the limit of very flat voids, coalescence can appear through diffuse plastic flow, as observed in both simulations \cite{thomasonnew2} and experiments (see Fig.~1). Estimating \textit{internal necking} coalescence load is thus still relevant in the limit of very flat voids $W \rightarrow 0$.}

\subsection{Theoretical expressions}
\label{theoexp}
Numerical simulations performed in \cite{thomasonnew2} have shown that, for flat voids, limit-load velocity field is not confined in the inter-void ligament but extends significantly above the voids. Therefore, the trial velocity field considered in the following (and shown on Fig.~1) extends up to a \textcolor{black}{constant} height $Z(r) = X \geq h$ in two separate regions denoted $\circled{1}$ and $\circled{2}$. In region $\circled{1}$, the trial velocity field - kinematically admissible (Eqs.~\ref{eqperiodic}) and verifying the property of incompressibility - proposed in \cite{thomasonnew2} is used:
\begin{equation}
\left\{
\begin{aligned}
v_r^{\circled{1}}(r,z) & = B(X-z)\left(\frac{L^2}{r}-r    \right) \\
v_z^{\circled{1}}(r,z) &= B(2Xz-z^{2})\\
\end{aligned}
\right.
\ \ \ \ \ \Rightarrow \ \ \ \ \  d_{eq}^{\circled{1}} = \frac{B}{\sqrt{3}} \sqrt{4(X-z)^{2} \left(\frac{L^4}{r^4}+3\right) + \left(\frac{L^2}{r}-r \right)^{2}   } 
\label{ve1}
\end{equation}
\textcolor{black}{where $B = [H D_{33}]/X^2$ (Eq.~\ref{eqperiodic})}. In region $\circled{2}$, the trial velocity field is chosen of the following form. The dependence in $r$ of the radial velocity ensures that the boundary condition at $r=0$ is satisfied:
\begin{equation}
\left\{
\begin{aligned}
v_r^{\circled{2}}(r,z) & = B(X-z) \, \alpha r \\
v_z^{\circled{2}}(r,z) &= g(z)\\
\end{aligned}
\right.
\label{ve2}
\end{equation}

\noindent where $g(z)$ is a function to be determined. To apply Eq.~\ref{eqanalyselimite} to assess the limit-load, trial velocity field should have only purely tangential discontinuity at any interface, which implies that:

\begin{equation}
v_r^{\circled{1}}(R,z) = v_r^{\circled{2}}(R,z) \ \ \ \ \ \Rightarrow \ \ \ \ \  \alpha = \frac{1-\chi^{2}}{\chi^{2}}
\end{equation}

\noindent
Verifying the property of incompressibility gives an equation for the unknown function $g(z)$:

\begin{equation}
\mathrm{tr}\ \textbf{d}^{\circled{2}} = \mathrm{tr} \begin{pmatrix}
\alpha B (X-z) & 0 & \displaystyle{-\frac{\alpha B r}{2}} \\
0 & \alpha B (X-z) & 0 \\
\displaystyle{-\frac{\alpha B r}{2}} & 0 & g'(z) \\
\end{pmatrix} = 0 \ \ \ \ \ \Rightarrow \ \ \ \ \ g'(z) + 2\alpha B (X-z) = 0
\label{eqincompressible}
\end{equation}

\noindent
Solving Eq.~\ref{eqincompressible} with the boundary condition $v_z^{\circled{1}}(r,X) = v_z^{\circled{2}}(r,X)$ imposed by the rigid part leads to:
\begin{equation}
\begin{aligned}
g(z) &= B[-\alpha(2Xz-z^{2}) + (1+\alpha)X^{2}   ]
\end{aligned}
\end{equation}
Finally, the equivalent strain rate in region $\circled{2}$ and the velocity jump at the interface between the two regions can be computed:
\begin{equation}
\begin{aligned}
d_{eq}^{\circled{2}} &= \frac{B \alpha}{\sqrt{3}} \sqrt{r^{2} + 12(X-z)^{2}}\\
||v_t||_{S_d} &= B(1+\alpha)(z-X)^{2}
\end{aligned}
\label{ve3}
\end{equation}

\noindent
The velocity field defined above is kinematically admissible with the macroscopic strain rate tensor \textbf{D} (Eqs.~\ref{eqperiodic}) chosen to assess coalescence, and verifies the property of incompressibility. It can therefore be used to assess the limit-load following equations~\ref{eqanalyselimite},~\ref{formulapi} and \ref{eqanalyselimite2}. We define the parameter $n$ such that:

\begin{equation}
n = \frac{X}{R}
\end{equation}

\noindent
The plastic zone thus extends up to an height $nR$ above the void. Taking $n=W$ will recover the continuous trial velocity field used in \cite{thomasonnew2}.\\

\noindent
The total macroscopic plastic dissipation is computed according to Eq.~\ref{formulapi} and is composed of three terms $\Pi = \Pi^1 + \Pi^2 + \Pi^3$. First term corresponds to the plastic dissipation in region $\circled{1}$ and is similar to the one computed in \cite{thomasonnew2}:

\begin{equation}
\begin{aligned}
\Pi^{1} &= \frac{1}{vol{\Omega}} \int_{\Omega_1} \sigma_0 d_{eq}^{\circled{1}} d\Omega = \frac{2\sigma_0 B}{\sqrt{3}L^2 H} \int_R^L rdr \int_0^{nR} \sqrt{4(X-z)^2\left(\frac{L^4}{r^4}+3   \right) + \left(\frac{L^2}{r} -r  \right)^2}dz\\
 &= \frac{L^2 \sigma_0 B}{\sqrt{3} H }  I_1(n,\chi)\\
 &\mathrm{with} \ \ \ \ \ I_1(n,\chi) = \int_{\chi^2}^{1} du  \int_0^{n \chi}  \frac{1-u}{\sqrt{u}}  \sqrt{1 + v^2 \frac{4(1+3u^2)}{u(1-u)^2}} dv 
\end{aligned}
\end{equation}

\noindent
$I_1$ can not be integrated analytically but can be transformed into a single integral, as shown in \cite{thomasonnew2} and in Appendix A. Its value can be computed using standard numerical integration procedure, the integrand behaving smoothly in any practical cases. The second term corresponds to the plastic dissipation in region $\circled{2}$:
\begin{equation}
\begin{aligned}
\Pi^{2} &= \frac{1}{vol{\Omega}} \int_{\Omega_2} \sigma_0 d_{eq}^{\circled{2}} d\Omega = \displaystyle{\frac{2B\sigma_0 \alpha}{\sqrt{3}  L^{2} H} \int_0^R  r dr  \int_h^{nR}  \sqrt{r^{2} + 12(X-z)^{2}   }   dz}\\
      &=  \frac{L^2 \sigma_0 B \alpha}{\sqrt{3} H } I_2(n,W,\chi)\\
 &\mathrm{with} \ \ \ \ \  I_2(n,W,\chi)     =  \int^{\chi^2}_{0} du  \int_0^{n\chi - W\chi}  \sqrt{u + 12v^{2}} dv
\end{aligned}
\end{equation}

\noindent
$I_2$ can be integrated analytically, the result is given in Appendix A. The last term corresponds to the tangential discontinuity at the interface between regions $\circled{1}$ and  $\circled{2}$:  

\begin{equation}
\begin{aligned}
\Pi^{3} &= \frac{1}{vol{\Omega}}  \int_{S_d} \frac{\sigma_0}{\sqrt{3}} ||v_t||_{S_d} dS = \displaystyle{   \frac{\sigma_0 B (1+\alpha)}{\sqrt{3} \pi L^{2} H} 2\pi R \int_{h}^{X} (z-X)^{2}dz    }\\     
       &= \displaystyle{\frac{2 B\sigma_0 (1+\alpha) R^{4}}{3\sqrt{3} L^{2} H} (n-W)^{3} }\\
\end{aligned}
\end{equation}

\noindent 
Finally, by noting that $D_{33} = [B R^{2} n^{2}]/H$ according to Eq.~\ref{eqperiodic}, Eq.~\ref{eqanalyselimite2} gives the equation of the upper-bound of the limit-load:
\begin{equation}
\frac{\Sigma_{33}}{\sigma_0} \leq \frac{I_1(n,\chi)}{\sqrt{3} n^{2} \chi^{2}} + \frac{(1-\chi^{2})}{\sqrt{3}n^{2} \chi^{4}}  I_2(n,W,\chi) + \frac{2}{3\sqrt{3}n^{2}}  (n-W)^{3}
\label{eqlimitload}
\end{equation}
The right-hand side of Eq.~\ref{eqlimitload} can be minimized with respect to the parameter $n$ (for $n\geq W$ which corresponds to $X \geq h$) to obtain the best estimate of the limit-load. It can be noticed that, for a given value of the parameter $n$, the limit-load remains finite when $W \rightarrow 0$. Setting $n = W$, the above equation reduces to the formula proposed in \cite{thomasonnew2}, which have been shown to be in very good agreement with results of numerical limit analysis for $W \geq 1$.\\

In order to use Eq.~\ref{eqlimitload} in the context of ductile-fracture modelling, a formula for the parameter $n$ as a function of $W$ and $\chi$ is required, to avoid performing the minimization procedure. The following choice:
\begin{equation}
\frac{\Sigma_{33}}{\sigma_0} \leq \frac{I_1(n_1,\chi)}{\sqrt{3} n_1^{2} \chi^{2}} + \frac{(1-\chi^{2})}{\sqrt{3}n_1^{2} \chi^{4}}  I_2(n_1,W,\chi) +  \frac{2}{3\sqrt{3}n_1^{2}}  (n_1-W)^{3} \ \ \ \ \ \mathrm{with}\ \ \ \ \ n_1 = \max{\left(\frac{1}{3\chi},W\right)}
\label{eqsimp}
\end{equation}
leads to predictions in very close agreement with the ones obtained by optimizing the parameter $n$, for $W \in [0:3]$ and $\chi \in [0.1:0.9]$. Eq.~\ref{eqsimp} will thus be used in the following instead of Eq.~\ref{eqlimitload}. One may note that $n = 1/[3\chi]$ corresponds to $X = L/3$, thus in most cases the extent of the \textcolor{black}{trial velocity field above the void that gives the best approximation of the real plastic dissipation} is on the order of the unit-cell width. The approximated value for the parameter $n$ also indicates that, whatever the aspect ratio of the void, \textcolor{black}{trial }velocity field (or plastic zone) extends above the void for sufficiently low value of $\chi \leq (3W)^{-1}$. \textcolor{black}{Such results do not not imply necessarily that the real velocity field in these situations extend well above the void height due to the crude approximation made for the trial velocity field of uniform vertical extension, which does not fulfill the conjecture proposed in \cite{benzerga02} that the rigid-plastic interface intercepts the pole of the cavity. This will be discussed in section~\ref{subseccomp} when comparing trial velocity field chosen here to the real velocity field obtained through numerical limit-analysis.} \\

Eq.~\ref{eqsimp} still relies on the computation of integral $I_1(n,W)$, but a closed-form formula for the upper-bound can be obtained using H\"older inequality by computing the plastic dissipation such as:\\
\begin{equation}
\begin{aligned}
\Pi^{1} &= \frac{2\sigma_0 B}{\sqrt{3}L^2 H} \int_R^L rdr \int_0^{nR} \sqrt{4(X-z)^2\left(\frac{L^4}{r^4}+3   \right) + \left(\frac{L^2}{r} -r  \right)^2}dz\\
       &\leq \frac{2\sigma_0 B}{\sqrt{3}L^2 H} \sqrt{nR(L-R)} \sqrt{\int_R^L dr  \int_0^{nR} 4 r^2 (X-z)^2\left(\frac{L^4}{r^4}+3   \right) + \left(L^2 -r^2  \right)^2dz}
\end{aligned}
\label{eqholder}
\end{equation}
Two cases must be distinguished: $W \geq 1/(3\chi)$ for which $n = W$ \textcolor{black}{and plastic dissipation reduces to $\Pi=\Pi^1$}, Eq.~\ref{eqholder} leads to:
\begin{equation}
\frac{\Sigma_{33}}{\sigma_0} \leq \frac{2\sqrt{W\chi(1-\chi)}}{3\sqrt{5}W^2 \chi^2} \sqrt{20W^3 (\chi^2 - \chi^6) - W(3\chi^6 - 10\chi^4 +15\chi^2 - 8\chi)} \ \ \ \ \ \mathrm{if}\ \ \ \ \ W \geq \frac{1}{3\chi}
\label{ll1}
\end{equation}
For $W \leq 1/(3\chi)$ for which $n = 1/[3\chi]$, the analytical expression of $I_2(n,W,\chi)$ is used (Appendix A), finally leading to:
\begin{equation}
\begin{aligned}
\frac{\Sigma_{33}}{\sigma_0} &\leq \frac{2 \sqrt{(1-\chi)}}{3\sqrt{5}\sqrt{\chi}} \sqrt{-27\chi^6 + 70\chi^4 - 135\chi^2 +72\chi +20}\\
&+  \frac{\sqrt{3}\chi^2(1-\chi^2)}{8}  \left[ \sqrt{3}\,\mathrm{asinh}\left( 2\,\sqrt{3}\,\beta\right) -96\sqrt{3}\,{\beta}^{4} +\sqrt{12\,{\beta}^{2}+1}\,\left( 48\,{\beta}^{3}+10\,\beta\right) \right]   \\
&+  2\sqrt{3}\chi^2 \beta^3 \ \ \ \ \ \mathrm{if}\ \ \ \ \ W \leq \frac{1}{3\chi}\\
&\mathrm{with}\ \ \ \ \ \beta = \frac{1}{3\chi} - W
\end{aligned}
\label{ll2}
\end{equation}
Predictions of both integral expression (Eq.~\ref{eqsimp}) and closed-form formula (Eqs.~\ref{ll1} and \ref{ll2}) for the coalescence limit-load of cylindrical voids by internal necking are now compared to the results of numerical limit analysis. 
\subsection{Comparisons to numerical limit analysis}
\label{subseccomp}
\begin{figure}[H]
\centering
\subfigure[]{\includegraphics[height = 5cm]{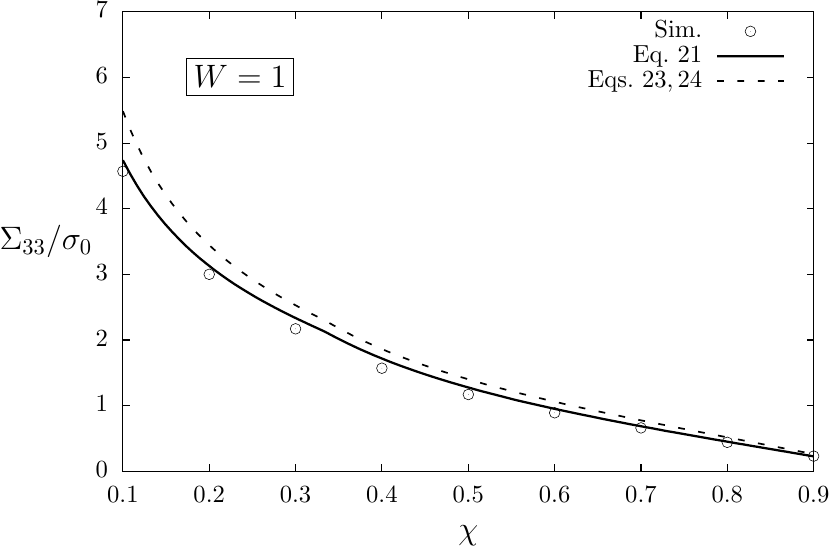}}
\subfigure[]{\includegraphics[height = 5cm]{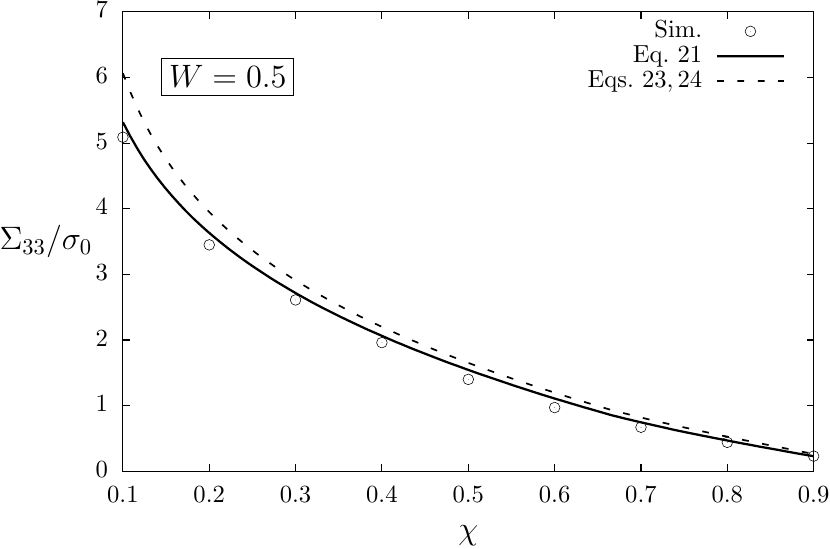}}
\subfigure[]{\includegraphics[height = 5cm]{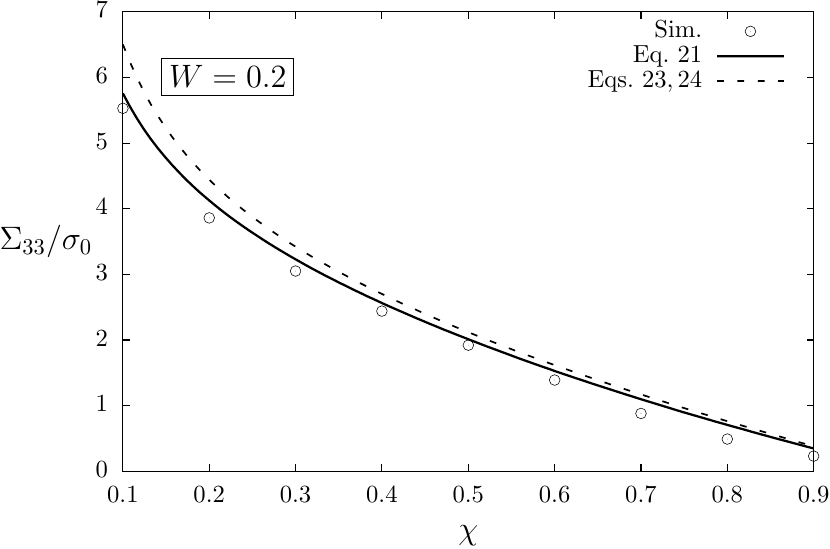}}
\subfigure[]{\includegraphics[height = 5cm]{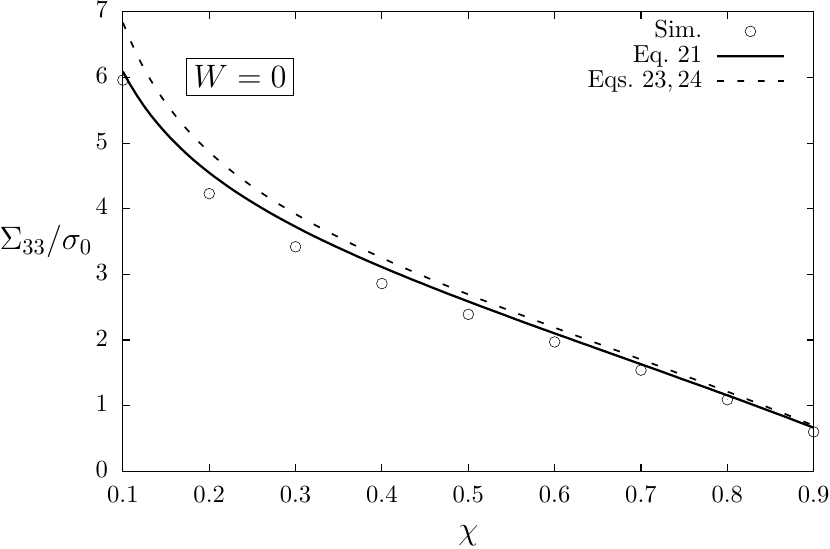}}%
\caption{Limit-load of cylindrical flat void in cylindrical unit-cell. Comparison of the theoretical upper-bound estimates to the results of numerical limit analysis as a function of the inter-void ligament dimensionless length, for different values of the void aspect ratio.}
\label{figcomp}
\end{figure}

\begin{figure}[H]
\centering
\subfigure[]{\includegraphics[height = 5cm]{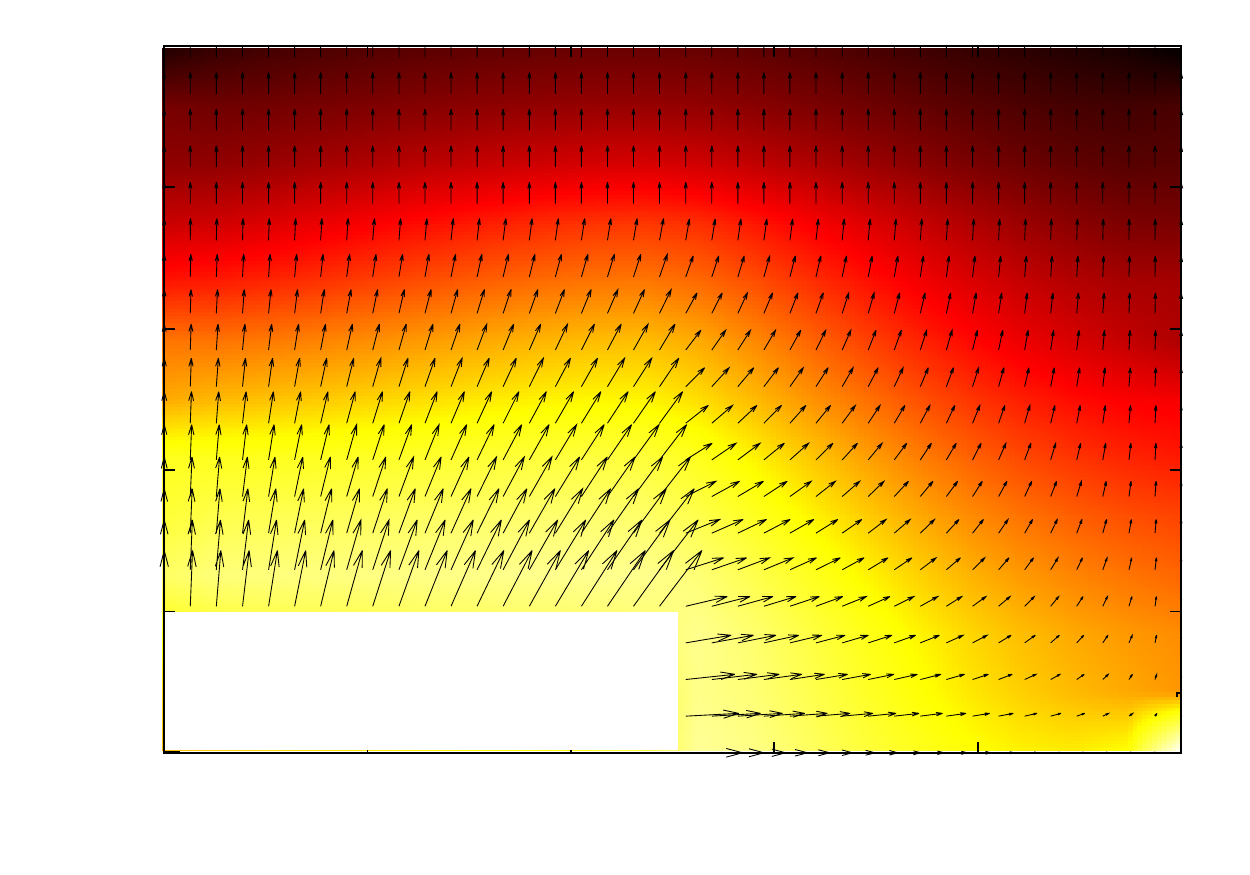}}
\subfigure[]{\includegraphics[height = 5cm]{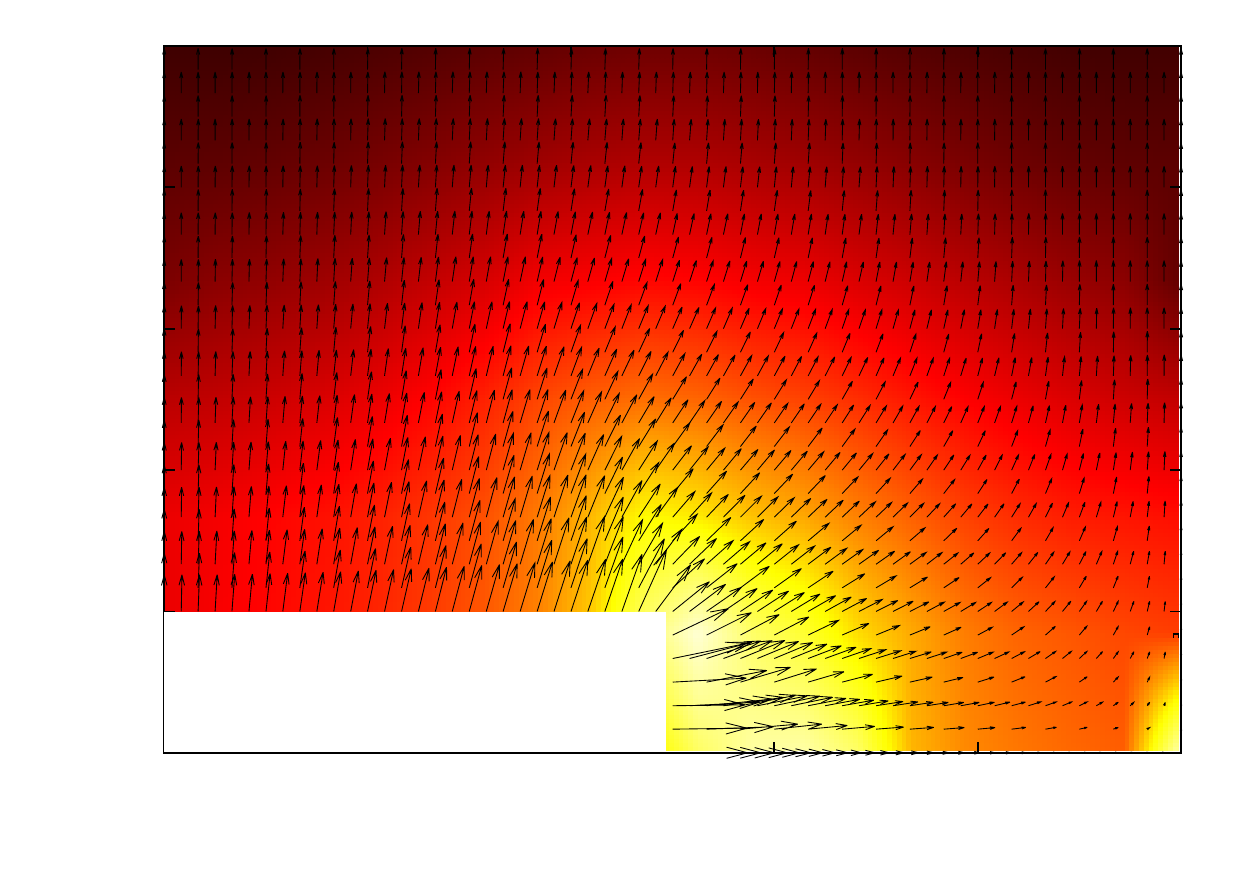}}
\caption{Velocity field (black arrows) and equivalent strain rate (colored map) for $W=0.2$ and $\chi=0.5$: (a) theoretical trial velocity field (b) numerical velocity field. As velocity field is defined up to a multiplicative factor, both velocity field have been rescaled for comparison purpose.}
\label{figcompcastem}
\end{figure}

Theoretical upper-bound estimates of the coalescence limit-load given by Eq.~\ref{eqsimp} and Eqs.~\ref{ll1},\ \ref{ll2} are plotted for different cylindrical flat void aspect ratio ($W=1,0.5,0.2,0$) and $\chi \in [0.1:0.9]$ in Fig.~\ref{figcomp}. Results are compared to the exact values (up to numerical errors) of the limit-load obtained through finite elements simulations (Section~\ref{num}). The integral expression of the upper-bound estimate of the limit-load (Eq.~\ref{eqsimp}) is found to be in very good agreement with numerical results for any aspect ratio, and in particular for penny-shaped cracks. Trial velocity field and associated equivalent strain rate are compared to the numerical results in Fig.~\ref{figcompcastem}. \textcolor{black}{The trial velocity field is able to capture the significant vertical extent of the plastic zone, explaining the good agreement between the upper bound estimate and the numerical limit-load. However, the real velocity field has an interface between the rigid and plastic region which is not planar. Improvement of the trial velocity field should consider an interface $X(r)$. As already mentionned in \cite{torki}, the conjecture that the interface intercepts the pole of the cavity \cite{benzerga02} may not be correct for very flat voids, as can be seen on Fig.~\ref{figcompcastem}b.} The closed-form formulas (Eqs.~\ref{ll1},\ref{ll2}) are found to be in rather good agreement with numerical results.

\begin{figure}[H]
\centering
\subfigure[]{\includegraphics[height = 5cm]{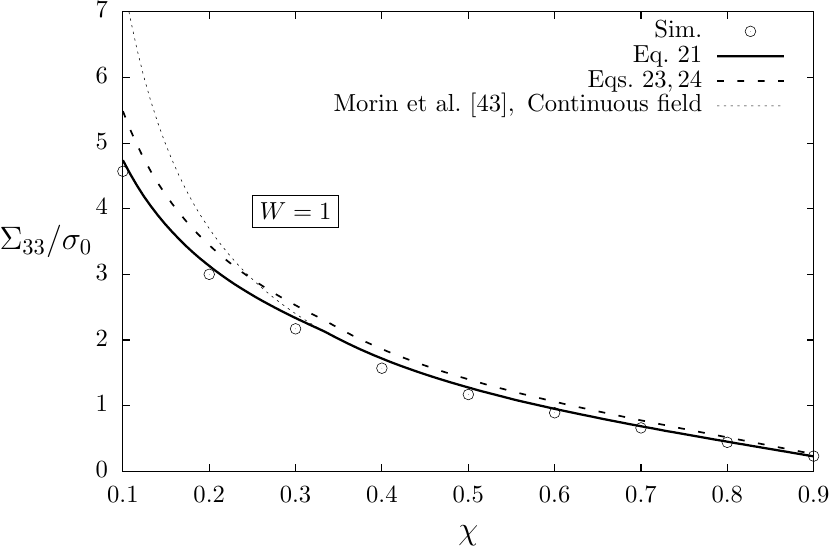}}
\subfigure[]{\includegraphics[height = 5cm]{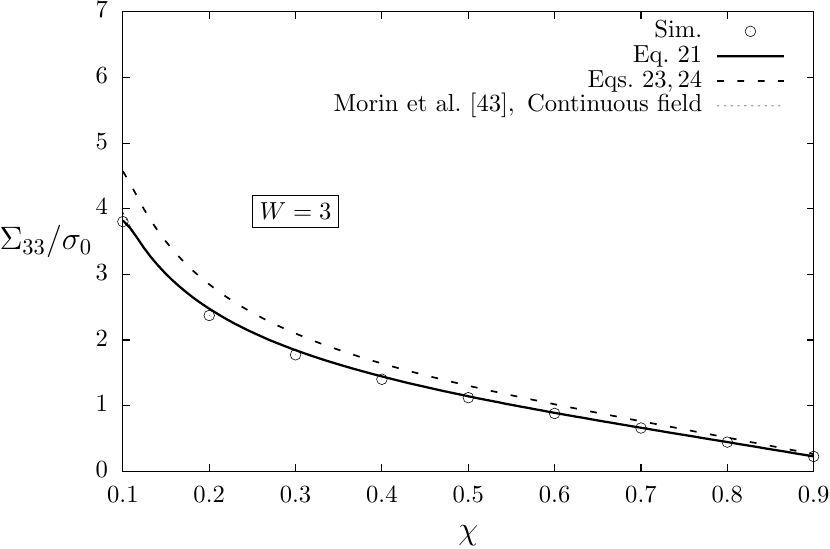}}
\caption{Limit-load of cylindrical elongated void in cylindrical unit-cell. Comparison of the theoretical upper-bound estimates to the results of numerical limit analysis and to the expression obtained in \cite{thomasonnew2} as a function of the inter-void ligament dimensionless length, for different values of the void aspect ratio.}
\label{figcompmorin}
\end{figure}

Eq.~\ref{eqsimp} is equivalent to the expression obtained in \cite{thomasonnew2} (and referred to as \textit{continuous field} hereafter) for elongated voids ($W \geq 1)$, except for very small value of $\chi$ where the plastic zone extends above the voids. Thus, estimates of Eq.~\ref{eqsimp} (and also Eq.~\ref{ll1},\ref{ll2}) are also compared to the expression obtained in \cite{thomasonnew2} for elongated voids in Fig.~\ref{figcompmorin}. \textcolor{black}{Although it might not be relevant for practical applications, it can be observed that} integral expression (Eq.~\ref{eqsimp}) (and closed-form formula Eqs.~\ref{ll1},\ref{ll2} to a lesser extent) improves the prediction of the limit-load obtained in \cite{thomasonnew2} for very small values of $\chi$, in the case of almost cubic void ($W \approx 1$)\footnote{Same conclusion also holds using the closed-form formula referred as \textit{Discontinuous field} of \cite{thomasonnew2}.}. The differences between the models disappear for elongated voids ($W \gg 1$).

\begin{figure}[H]
\centering
\subfigure[]{\includegraphics[height = 5cm]{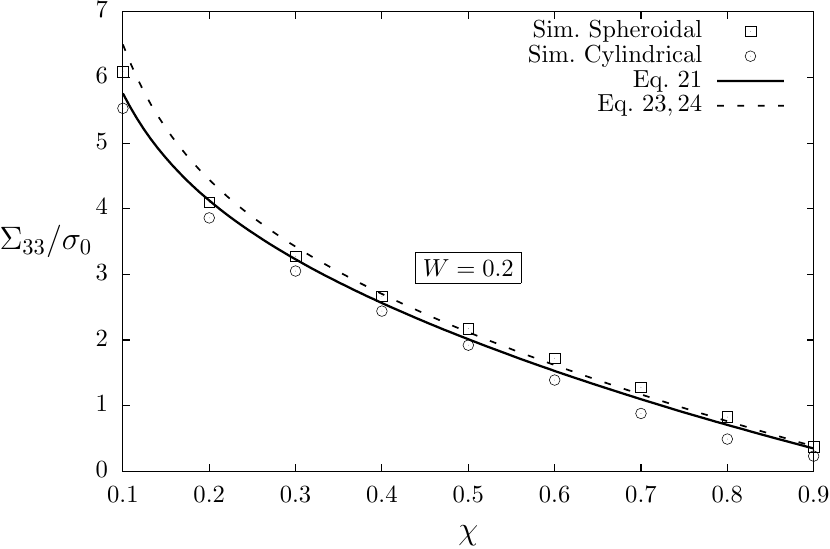}}
\subfigure[]{\includegraphics[height = 5cm]{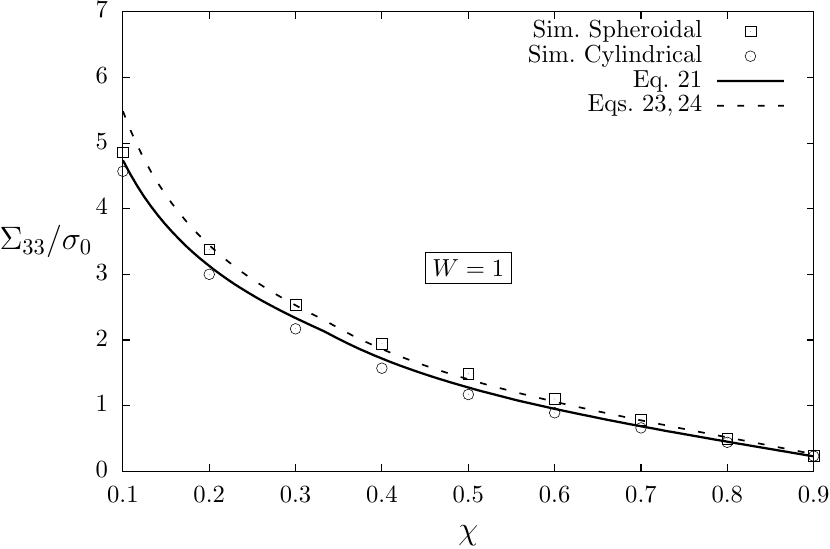}}
\caption{Limit load of void in cylindrical unit-cell. Comparison of the theoretical upper-bound estimates to the results of numerical limit analysis for both cylindrical and spheroidal voids as a function of the inter-void ligament dimensionless length, for different values of the void aspect ratio.}
\label{figcompsphere}
\end{figure}

Eq.~\ref{eqsimp} and Eqs.~\ref{ll1},\ref{ll2} are finally compared to numerical results obtained with spheroidal voids in Fig.~\ref{figcompsphere}. A good agreement is observed for flat ($W \ll 1$), spherical ($W \approx 1$) and elongated ($W \gg 1$, not shown here) voids, justifying these upper-bound estimates can be used to predict coalescence for spheroidal voids. One may note however that in this case, the theoretical estimates are no longer upper-bounds. To be complete, it can be mentioned that the choice of the cylindrical unit-cell influences only weakly the limit-load obtained compared to cubic unit-cell considering a given porosity, as shown in \cite{torki}.

\subsection{\textcolor{black}{Extension to combined tension and shear}}

\begin{figure}[H]
\centering
\includegraphics[height = 5cm]{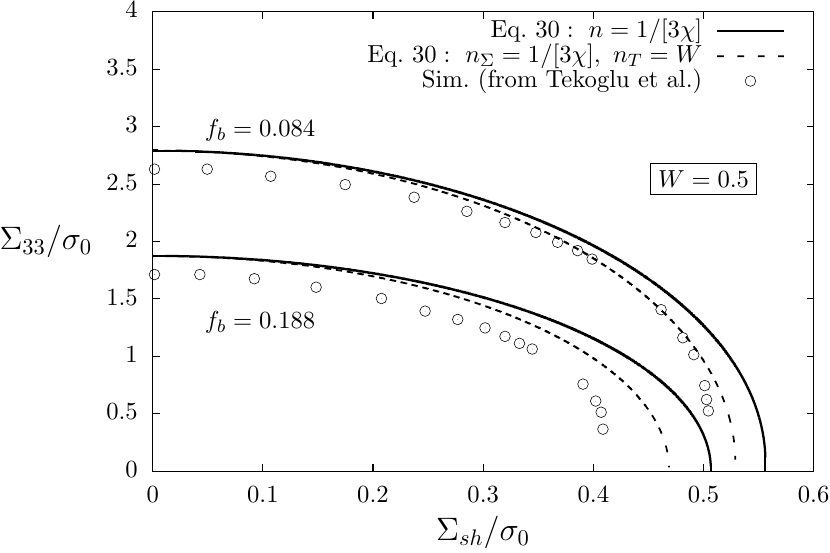}
\caption{\textcolor{black}{Comparison of the theoretical expressions (Eq.~\ref{eqfullshear}) and numerical results taken from \cite{tekoglushear} for the value of the coalescence stress under combined tension and shear loading ($\Sigma_{sh}^2 = \Sigma_{31}^2 + \Sigma_{32}^2$) for two different values of the porosity of the layer containing the void.}}
\label{compshear}
\end{figure}

\textcolor{black}{The coalescence criterion derived in the previous section can be extended to account for combined tension and shear following the work of Torki \textit{et al.} \cite{torki}. Their analysis is reproduced hereafter and adapted to the trial velocity field considered here:}
\begin{equation}
\textcolor{black}{v = v^E + v^S}
\end{equation}
\textcolor{black}{where $v^E$ is an extensional velocity field induced by tension and taken equal to Eqs.~\ref{ve1}, \ref{ve2}, and $v^S$ is induced by shear:}
\begin{equation}
\left\{
\begin{aligned}
\textcolor{black}{v_r^S(r,z)} &  \textcolor{black}{=\frac{2zH}{X}(D_{31}\cos{\theta} + D_{32}\sin{\theta}    )} \\
\textcolor{black}{v_{\theta}^S(r,z)} & \textcolor{black}{=\frac{2zH}{X}(-D_{31}\sin{\theta} + D_{32}\cos{\theta}    )}\\
\end{aligned}
\right.
\end{equation}
\textcolor{black}{in both regions $\circled{1}$ and $\circled{2}$. The trial velocity field is kinematically admissible with $\bm{D}=D_{33} e_3\otimes e_3 + D_{31} (e_1\otimes e_3 + e_3\otimes e_1) + D_{32}  (e_2\otimes e_3+ e_3\otimes e_2)$, and verifies the property of incompressibility. Due to the orthogonality of the tension and shear induced strain rate, the equivalent strain rate can be computed as:}
\begin{equation}
\begin{aligned}
\textcolor{black}{d_{eq}^2} &  \textcolor{black}{= \frac{2}{3}\left(\bm{d}^E:\bm{d}^E + \bm{d}^S:\bm{d}^S\right)} \\
                        &  \textcolor{black}{=\left(d_{eq}^E\right)^2 + \frac{4}{3} \frac{H^2}{X^2} \left(D_{31}^2  + D_{32}^2\right)     }
\end{aligned} 
\end{equation}
\textcolor{black}{where $d_{eq}^E$ corresponds to Eqs.~\ref{ve1}, \ref{ve3} in region $\circled{1}$ and $\circled{2}$, respectively. Plastic dissipation can be estimated as:}
\begin{equation}
\begin{aligned}
\textcolor{black}{\Pi} & \textcolor{black}{\leq \left< \sigma_0 d_{eq} \right>_{\Omega - \omega}} \textcolor{black}{+ \frac{1}{vol{\Omega}} \int_{Sd} \frac{\sigma_0}{\sqrt{3}} \| v_t\|dS}\\
                     & \textcolor{black}{\approx \sigma_0  \sqrt{ \left< d_{eq}^E \right>_{\Omega - \omega}^2 + \left< \frac{2}{\sqrt{3}} \frac{H}{X} \sqrt{D_{31}^2  + D_{32}^2}  \right>_{\Omega - \omega}^2 }  + \displaystyle{\frac{2 B\sigma_0 (1+\alpha) R^{4}}{3\sqrt{3} L^{2} H} (n-W)^{3} }      } \\ 
              &  \textcolor{black}{ \approx \sqrt{\Sigma^{vol^2} D_{33}^2   + T^2 (D_{31}^2  + D_{32}^2) }  + \Sigma^{surf}|D_{33}|}
\end{aligned}
\end{equation}
\textcolor{black}{where the approximation made to estimate the plastic dissipation may not be upper-bound preserving \cite{tekoglushear,torki}. $\Sigma^{vol}$, $\Sigma^{surf}$ and $T$ are defined as:}
\begin{equation}
\begin{aligned}
\textcolor{black}{\Sigma^{vol}} & \textcolor{black}{=\sigma_0 \left[ \frac{I_1(n,\chi)}{\sqrt{3} n^{2} \chi^{2}} + \frac{(1-\chi^{2})}{\sqrt{3}n^{2} \chi^{4}}  I_2(n,W,\chi) \right]}\\
\textcolor{black}{\Sigma^{surf}} & \textcolor{black}{=\frac{2 \sigma_0}{3\sqrt{3}n^{2}}  (n-W)^{3}}\\
\textcolor{black}{T} & \textcolor{black}{=\frac{2 \sigma_0}{\sqrt{3}} \left(1 - \frac{W\chi^2}{n}    \right)}
\end{aligned}
\end{equation}
\textcolor{black}{The plastic dissipation is in the same form as the one given in \cite{torki}, where it has been shown that the void coalescence criterion reads:}
\begin{equation}
\left\{  
\begin{aligned}
\textcolor{black}{\frac{(|\Sigma_{33}| - \Sigma^{surf} )^2}{\Sigma^{vol^2}} + 4 \frac{\Sigma_{31}^2 + \Sigma_{32}^2}{T^2} - 1} & \textcolor{black}{= 0} & \textcolor{black}{\mathrm{for}\ |\Sigma_{33}| \geq \Sigma^{surf}}\\ 
\textcolor{black}{4 \frac{\Sigma_{31}^2 + \Sigma_{32}^2}{T^2} - 1} & \textcolor{black}{= 0} & \textcolor{black}{\mathrm{for}\ |\Sigma_{33}| \leq \Sigma^{surf}}\\ 
\end{aligned}
\right.
\label{eqfullshear}
\end{equation}
\textcolor{black}{The coalescence criterion (Eq.~\ref{eqfullshear}) is compared to the numerical results of Teko{\u{g}}lu \textit{et al.} \cite{tekoglushear}. As cylindrical unit-cell with cylindrical voids are considered here while simulations were performed with cubic unit-cell and spherical voids, comparisons are made at constant porosity in the porous band $f_b$, where $f_b=\chi^2$ in former case, and $f_b=[\pi/6]\chi^2$ in the latter, as in \cite{torki,keralavarma}. In principle, the parameter $n$ should be optimized so as to find the minimum plastic dissipation. However, for simplicity, two cases are considered here: in the first one, its value is taken equal to the approximate value obtained for axisymmetric loading conditions in section~\ref{theoexp} $n=\max{(1/[3\chi],W)}$ while in the second, the approximate value is taken only in the definition of $\Sigma^{vol}$ and $\Sigma^{surf}$, while $n=W$ is taken for $T$. This second case leads to optimal values in both pure shear and axisymmetric tension. Results are shown for $W=0.5$ and two values of porosity in Fig.~\ref{compshear}. Good agreement is observed between theoretical predictions and numerical results. Despite the approximation made to compute the plastic dissipation and regarding the choice of parameter $n$, theoretical expressions are found to be upper-bounds. The choice of different values of $n$ for $\Sigma^{vol}$,  $\Sigma^{surf}$ and $T$ leads to predictions in close agreement with numerical results, although coalescence load is overestimated for pure shear. This is believed to come from the rather crude approximation of the shear component of the velocity field.}

\section{Discussion and Conclusions}

\textcolor{black}{For isotropic porous material,} in order to overcome the drawback of the Thomason coalescence model predicting infinite coalescence load for penny-shaped cracks, Benzerga \cite{benzerga02} proposed an empirical modification \textcolor{black}{of Thomason equation:}
\begin{equation}
\textcolor{black}{\frac{\Sigma_{33}}{\sigma_0} = (1 - \chi^2) \left[0.1 \left[\frac{\chi^{-1} - 1}{f(W,\chi)}     \right]^2 + A \sqrt{\chi^{-1}}       \right]}
\label{eqbenzerga}
\end{equation}
\textcolor{black}{where the original formulas for $f(W,\chi)=W$ and $ A=1.2$ have been replaced by $f(W,\chi)=W^2 + 0.1 \chi^{-1} + 0.02 \chi^{-2}$ and $A=1.3$.} \textcolor{black}{Torki \textit{et al.} \cite{torki} proposed an heuristic extension of the rigorous upper-bound criterion for coalescence developped in \cite{thomasonnew1}:}
\begin{equation}
\textcolor{black}{\frac{\Sigma_{33}}{\sigma_0} = t(W,\chi) \left[\frac{\chi^3 - 3\chi + 2}{3\sqrt{3} W\chi} \right]     + \frac{b}{\sqrt{3}}\left[2 - \sqrt{1+3\chi^4} + \ln \frac{1 + \sqrt{1 + 3\chi^4}}{3\chi^2}    \right]}
\label{eqtorki}
\end{equation}
\textcolor{black}{where the original formulas for $t(W,\chi)=1$ and $ b=1$ have been replaced by $t(W,\chi)=W[-0.84 + 12.9\chi]/[1 + W(-0.84 + 12.9\chi)]$ and $b=0.9$.}
\textcolor{black}{In a similar approach, Keralavarma and Chockalingam \cite{keralavarma} proposed the following expression:}
\begin{equation}
\textcolor{black}{\frac{\Sigma_{33}}{\sigma_0} = \sqrt{\frac{6}{5}} \left[b \ln\frac{1}{\chi^2} + \sqrt{b^2+1} - \sqrt{b^2 + \chi^4} + b \ln\left(\frac{b + \sqrt{b^2 + \chi^4}}{b + \sqrt{b^2 + 1}}    \right)                \right]}
\label{eqkeralavarma}
\end{equation}
\textcolor{black}{with $b^2 = 1/3 + [5 \alpha]/[24 \bar{W}^2]$, $\alpha = [1 + \chi^2 - 5\chi^4 + 3\chi^6]/12$ and $\bar{W} = W\chi\ \mathrm{ if }\ W\chi \geq 2W_0\ \mathrm{, }=W_0 + [W\chi]^2/[4W_0]\ \mathrm{ otherwise}$, with $W_0=0.12.$}
 
\begin{figure}[H]
\centering
\subfigure[]{\includegraphics[height = 5cm]{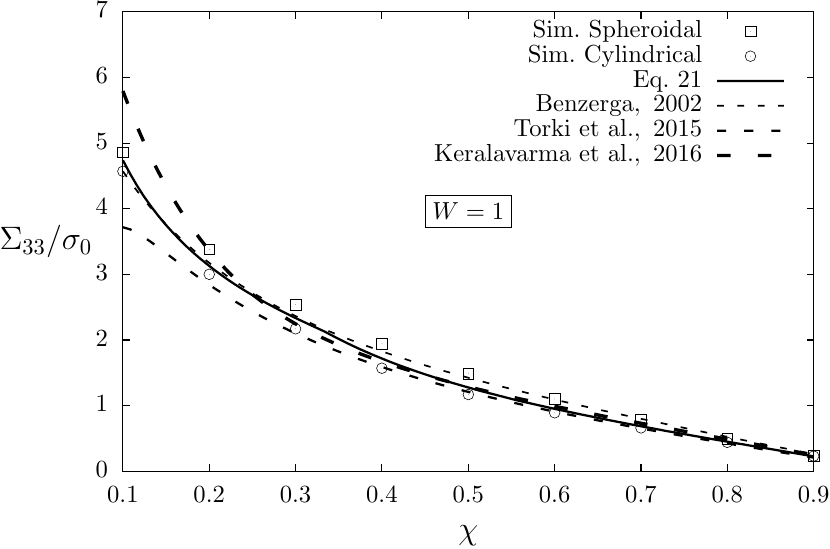}}
\subfigure[]{\includegraphics[height = 5cm]{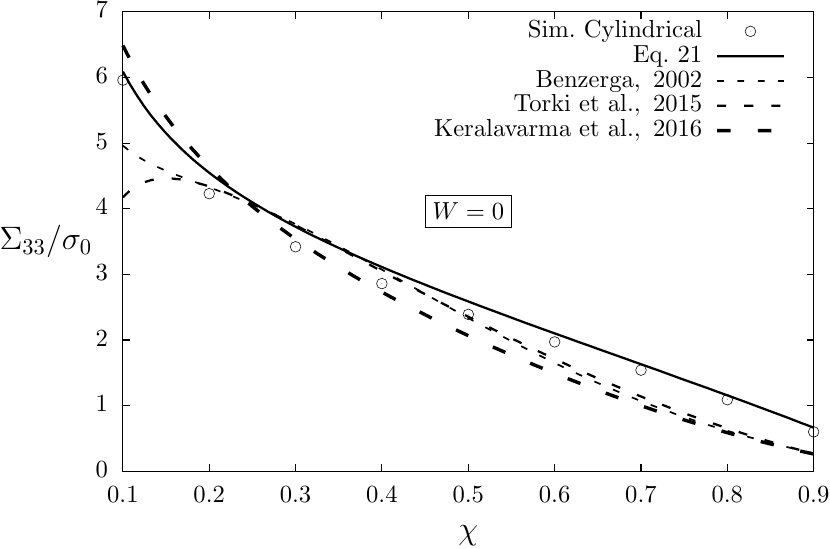}}
\caption{Limit-load of void in cylindrical unit-cell. Comparison of the theoretical upper-bound estimate (Eq.~21), Benzerga model (Eq.~\ref{eqbenzerga}) \cite{benzerga02}, Torki \textit{et al.} model (Eq.~\ref{eqtorki}) \cite{torki} \textcolor{black}{and Keralavarma and Chockalingam model (Eq.~\ref{eqkeralavarma}) \cite{keralavarma}} to the results of numerical limit analysis for both cylindrical and spheroidal voids as a function of the inter-void ligament dimensionless length, for different values of the void aspect ratio.}
\label{compbenzerga}
\end{figure}

These \textcolor{black}{three} models are compared to the integral expression for the upper-bound estimate of the limit-load obtained in this study for cylindrical voids in Fig.~\ref{compbenzerga}. All models give rather good estimates of the limit-load for both cylindrical and spheroidal flat voids ($0 \leq W \leq 1$) and can be used for practical purpose. Eq.~21 seems however preferable for very flat voids where it gives both a rigorous upper-bound estimate and better predictions for very \textcolor{black}{high values of the intervoid ligament dimensionless length, that might be relevant for practical applications involving coalescence of penny-shaped cracks}.

To conclude, based on limit analysis, rigorous theoretical upper-bound estimates - both integral expression and closed-form formula - of the limit-load of a cylindrical void in a cylindrical unit-cell with boundary conditions compatible with coalescence have been proposed. These upper-bound estimates do not exhibit singular behavior when the aspect ratio of the void goes to zero and are shown to be in very good agreement with exact limit-load obtained with finite elements simulations, for both cylindrical and spheroidal voids, over large ranges of parameters $W \in [0:3]$ and $\chi \in [0.1:0.9]$. The results obtained in this study thus improve the ones obtained in \cite{thomasonnew2} for flat voids, and those of \cite{benzerga02,torki,keralavarma} for penny-shaped cracks ($W \rightarrow 0$). \textcolor{black}{The coalescence criterion has been extended to account for combined tension and shear following the methodology proposed in \cite{torki}, and predictions are in good agreement with numerical results.} \textcolor{black}{The trial velocity field considered here leads to analytical limit-loads in close agreement with those obtained through numerical limit-analysis. However, one can note that the assumption of planar rigid-plastic interface and the presence of a velocity discontinuity are contrary to what is observed for the former, and unphysical for the latter. Future work should focus on finding a refined continuous trial velocity field with a non-planar rigid-plastic interface as observed in simulations.} \textcolor{black}{The proposed theoretical expressions (Eqs.~\ref{eqsimp}, \ref{eqfullshear}) can be used to model the post-coalescence effective behavior of isotropic porous material as described in details in \cite{benzerga02,benzergaleblond}. In this approach, Eqs.~\ref{eqsimp}, \ref{eqfullshear} are considered as the yield surfaces, and are supplemented by the evolution of void and ligament dimensions \cite{benzerga02} and accounting for strain-hardening \cite{benzergaleblond}.}

\section{Appendix A}

The integral expressions introduced in Eq.~\ref{eqlimitload} can be integrated analytically for $I_2(n,W,\chi)$ and reduced to a 1D integral for $I_1(n,\chi)$. In any practical cases ($\chi > 0$), the integrand behaves smoothly and the integral can be easily evaluated numerically with standard integration procedure. 
\begin{equation}
\begin{aligned}
I_1(n,\chi) &= \int_{\chi^2}^{1} du  \int_0^{n \chi}  \frac{1-u}{\sqrt{u}}  \sqrt{1 + v^2 \frac{4(1+3u^2)}{u(1-u)^2}} dv \\
     &= \int_{{\chi}^{2}}^{1} \left[   -\frac{(x-1)^2}{4\sqrt{1+3x^2}}\mathrm{asinh}\left(\frac{2\chi n \sqrt{1+3x^2}}{(x-1)\sqrt{x}}\right)  + \frac{\chi n}{2x}\sqrt{x^3 + 12\chi^2 n^2 x^2 - 2x^2 + x + 4\chi^2 n^2}     \right]dx
\end{aligned}
\end{equation}

\begin{equation}
\begin{aligned}
 I_2(n,W,\chi)     &=  \int^{\chi^2}_{0} du  \int_0^{n\chi - W\chi = \chi \alpha}  \sqrt{u + 12v^{2}} dv \\
                   &=\frac{\left( \sqrt{3}\,\mathrm{asinh}\left( 2\,\sqrt{3}\,\alpha\right) -32\,{3}^{\frac{3}{2}}\,{\alpha}^{4}\right) \,{\chi}^{4}+\sqrt{12\,{\alpha}^{2}+1}\,\left( 48\,{\alpha}^{3}+10\,\alpha\right) \,{\chi}^{4}}{24}
\end{aligned}
\end{equation}

\noindent
\textbf{Acknowlegments}\\

J.H. thanks J.B. Leblond for fruitful discussions. The authors thank B. Tanguy for comments on the manuscript.

\bibliographystyle{elsarticle-num.bst}
\bibliography{spebib2}

\end{document}